\documentclass[journal]{IEEEtran}
\usepackage{cite}
\usepackage[numbers,sort&compress]{natbib}
\usepackage{amsmath,bm}
\usepackage{threeparttable}
\usepackage{cite}

\ifCLASSINFOpdf
  \usepackage[pdftex]{graphicx}
\else
\fi
\usepackage{amsmath}
\usepackage{amssymb}
\usepackage{array}
\usepackage[caption=false,font=footnotesize]{subfig}
\usepackage{makecell}

\begin{document}
%
\title{Non-Integer-Oversampling Digital Signal Processing for Coherent Passive Optical Networks}

\author{Haide~Wang,
        Ji~Zhou,
        Jinyang~Yang,
        Jianrui~Zeng,
        Weiping~Liu,
        Changyuan~Yu,
        Fan~Li,
        and~Zhaohui~Li

\thanks{This work was supported in part by the National Natural Science Foundation of China (62005102); Hong Kong Scholars Program (XJ2021018). (Corresponding authors: Ji Zhou, e-mail: zhouji@jnu.edu.cn), (H. Wang and J. Yang contributed equally to this work).}

\thanks{H. Wang, J. Zhou, J. Yang, J. Zeng, and W. Liu are with the Department of Electronic Engineering, College of Information Science and Technology, Jinan University, Guangzhou 510632, China. }

\thanks{J. Zhou and C. Yu are with the Department of Electronic and Information Engineering, The Hong Kong Polytechnic University, Hong Kong.}

\thanks{F. Li and Z. Li are with the Guangdong Provincial Key Laboratory of Optoelectronic Information Processing Chips and Systems, Sun Yat-sen University, Guangzhou 510275, China and Southern Marine Science and Engineering Guangdong Laboratory (Zhuhai), Zhuhai, China.}}
%
%

\markboth{}%
{Shell \MakeLowercase{\textit{et al.}}: Bare Demo of IEEEtran.cls for IEEE Journals}
%



\maketitle

\begin{abstract}
Beyond 100G passive optical networks (PONs) will be required to meet the ever-increasing traffic demand in the future. Coherent optical technologies are the competitive solutions for the future beyond 100G PON but also face challenges such as the high computational complexity of digital signal processing (DSP). A high oversampling rate in coherent optical technologies results in the high computational complexity of DSP. Therefore, DSP running in a non-integer-oversampling below 2 samples-per-symbol (sps) is preferred, which can not only reduce computational complexity but also obviously lower the requirement for the analog-to-digital converter. In this paper, we propose a non-integer-oversampling DSP for meeting the requirements of coherent PON. The proposed DSP working at 9/8-sps and 5/4-sps oversampling rates can be reduced by 44.04\% and 40.78\% computational complexity compared to that working at the 2-sps oversampling rate, respectively. Moreover, a 400-Gb/s-net-rate coherent PON based on digital subcarrier multiplexing was demonstrated to verify the feasibility of the non-integer-oversampling DSP. There is almost no penalty on the receiver sensitivity when the non-integer-oversampling DSP is adopted. In conclusion, the non-integer-oversampling DSP shows great potential in the future coherent PON.
\end{abstract}

\IEEEpeerreviewmaketitle

\section{Introduction}
\IEEEPARstart{I}{t} is foreseeable that the beyond 100G passive optical networks (PONs) will be required to meet the ever-increasing traffic demand in the future \cite{faruk2021coherent, zhou2022100g}. Coherent optical technologies are currently moving towards the access network from core and metro networks, which are the competitive solutions for the future beyond 100G PON \cite{zhang2022coherent, campos2023coherent}. However, coherent optical technologies have a high cost for the future 100G PON \cite{zhou2021non}. Simplified coherent optical technologies are widely studied, such as a single-polarization heterodyne receiver and the single Mach-Zehnder modulator at the optical network unit (ONU) rather than the full coherent transceivers \cite{xing2023first}. Furthermore, digital subcarrier multiplexing (DSCM) is another appealing solution to implement coherent PON in an efficient and economical manner \cite{zhang2020rate, xu2022intelligent}. Coherent DSCM technology allows the low-bandwidth transceivers at ONU, which is preferred for the cost-sensitive PON scenario.

\begin{figure*}[!t]
\centering
\includegraphics[width=\linewidth]{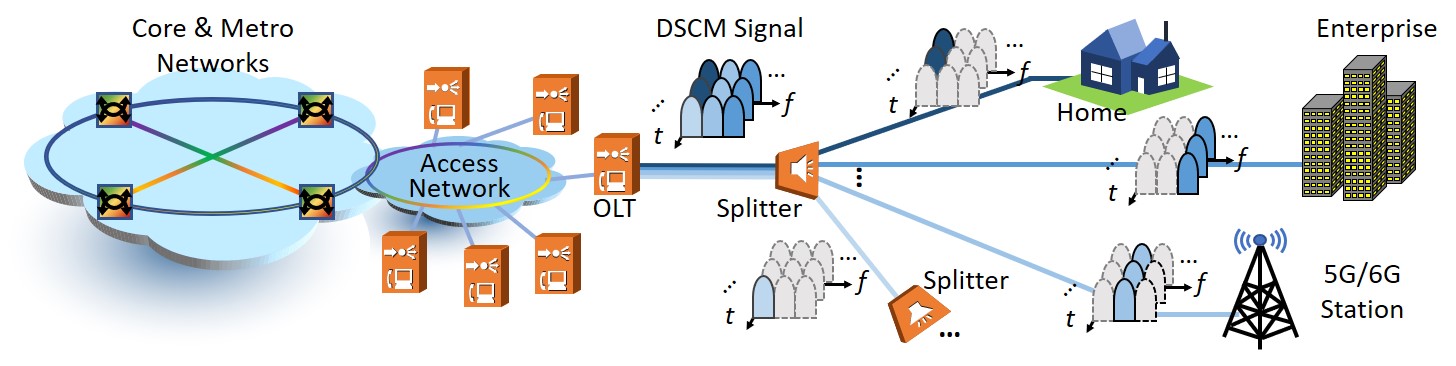}
\caption{Diagram of the future TFDMA-coherent PON using DSCM, where the coherent optical technologies are moving from the core and metro networks towards the access network. The allocated subcarriers of the DSCM signal can be selected by ONUs using the low-bandwidth transceivers.}
\label{NETWORK}
\end{figure*}

Coherent PON using DSCM can support massive and low-latency connectivity by providing time-and-frequency division multiple access (TFDMA). A diagram of the future TFDMA-coherent PON is shown in Fig. \ref{NETWORK}, where the allocated subcarriers of the DSCM signal from the optical line terminal (OLT) can be selected by ONUs. Therefore, the burden of high cost for TFDMA-coherent PON would be alleviated by using the low-bandwidth transceivers. However, the high computational complexity of digital signal processing (DSP) is still a great challenge for TFDMA-coherent PON \cite{nagarajan2021low}. To reduce the high computational complexity, simplified DSP is an important solution \cite{zhang2022intra, song2023low}. The conventional adaptive equalizer can be split into a 1-tap butterfly finite impulse response (FIR) filter and two non-butterfly FIR filters to halve the tap number of equalizers \cite{matsuda2018hardware}. In a simplified carrier phase recovery (CPR), the phase estimation of the X polarization can be shared with the Y polarization, which removes the phase estimation of the Y polarization \cite{suzuki2022simplified}. 

A high oversampling rate in coherent optical technologies increases the number of samples to be processed, which is another factor that results in the high computational complexity of DSP \cite{kaneda2008coherent, duthel2007impairment, savory2010digital}. Therefore, the non-integer-oversampling DSP is another effective solution for coherent PON to reduce computational complexity. The non-integer-oversampling DSP requires an oversampling rate below two samples-per-symbol (sps) \cite{malouin2010sub, baeuerle2018low, arikawa2023frequency}, which directly lowers the computational complexity by reducing the total number of processing samples. Moreover, the non-integer-oversampling DSP can lower the high sampling rate requirement of the analog-to-digital converter (ADC), which would also reduce the cost and is highly preferred for the ONUs in coherent PON. However, the non-integer-oversampling DSP for TFDMA-coherent PON using DSCM still lacks adequate research.

In this paper, we propose a non-integer-oversampling DSP to reduce the computational complexity for TFDMA-coherent PON using DSCM. The non-integer-oversampling DSP works at an oversampling rate close to 1 sps, which reduces the computational complexity of DSP as well as the high sampling rate requirement of the ADC. The main contributions of the work are as follows:
\begin{itemize}
\item The non-integer-oversampling DSP is proposed for coherent PON, which mainly includes the non-integer-oversampling frequency offset estimation (FOE), timing recovery (TR), frame synchronization, and polyphase fractionally-spaced multiple-input-multiple-output (MIMO) equalizer. 
\item  A coherent PON using DSCM is demonstrated to verify the feasibility of the non-integer-oversampling DSP. Compared to the DSP at a 2-sps oversampling rate, the computational complexity of the proposed DSP at 9/8-sps and 5/4-sps oversampling rates can be reduced by 44.04\% and 40.78\% almost without penalty on receiver sensitivity, respectively.
\end{itemize}

The remainder of this paper is organized as follows. In Section \ref{principle}, the principle of non-integer-oversampling DSP is given, which includes the non-integer-oversampling coarse FOE, TR, frame synchronization and fine FOE, polyphase fractionally-spaced MIMO equalizer. In Section \ref{experimental_setups}, the experimental setups of an $8$-Gbaud/subcarrier (SC)$\times$8-SCs DSCM-based coherent PON are introduced, which adopts the proposed non-integer-oversampling DSP. The experimental results and discussions are given in Section \ref{ex_results}. Finally, the paper is concluded in Section \ref{conclusions}.

\section{Principle of non-integer-oversampling DSP}\label{principle}
In this section, the principle of the proposed non-integer-oversampling DSP will be introduced, including the coarse FOE, TR, frame synchronization, fine FOE, MIMO equalizer, and CPR. The non-integer-oversampling coarse FOE and TR use the frequency-domain (FD) characteristics of the training sequence to solve the frequency offset and timing error. Since frame synchronization and fine FOE rely on the time-domain (TD) characteristics of the training sequence, the subsequent DSP is executed in the TD. Therefore, the FD and TD algorithms are combined in non-integer-oversampling DSP. 

\subsection{Non-integer-oversampling coarse FOE}
The coarse FOE is implemented using a training sequence with the two symmetric frequency tones at $\pm 1/2T$, where $T$ is the symbol period. If there is the frequency offset $\Delta f$, the two frequency tones will locate at $1/2T + \Delta f$ and $-1/2T + \Delta f$. Therefore, the coarse FOE is the average of the two tones. The coarse frequency offset is recovered by a cyclic shift of the frequency components, which is transformed by discrete Fourier transform (DFT). The FD algorithms are implemented using the overlap-save method with an overlap rate $\eta$ $(\eta > 1)$.

When the signal is oversampled at a $K/M$-sps oversampling rate, the accuracy of the coarse FOE can be calculated as
\begin{equation}
accuracy = \frac{K}{M N T}
\end{equation}
where $N$ is the DFT size and $M < K$. Although a lower sampling rate enables more accurate FOE, the high-frequency components of one sideband would be filtered due to the frequency offset. For the Nyquist-shaped signal with a roll-off factor of $\beta$, the minimum oversampling rate without performance penalty can be calculated as
\begin{equation}
K/M_{min} = 1 + \beta .
\end{equation}
Therefore, the higher the sampling rate, the greater the frequency offset the signal can tolerate. Although increasing the size of DFT makes FOE more accurate, it also makes computational complexity higher. Thus, a small-size DFT-based FOE requires a fine estimation of the residual frequency offset.

\subsection{Non-integer-oversampling timing recovery}
\begin{figure}[!t]
\centering
\includegraphics[width=\linewidth]{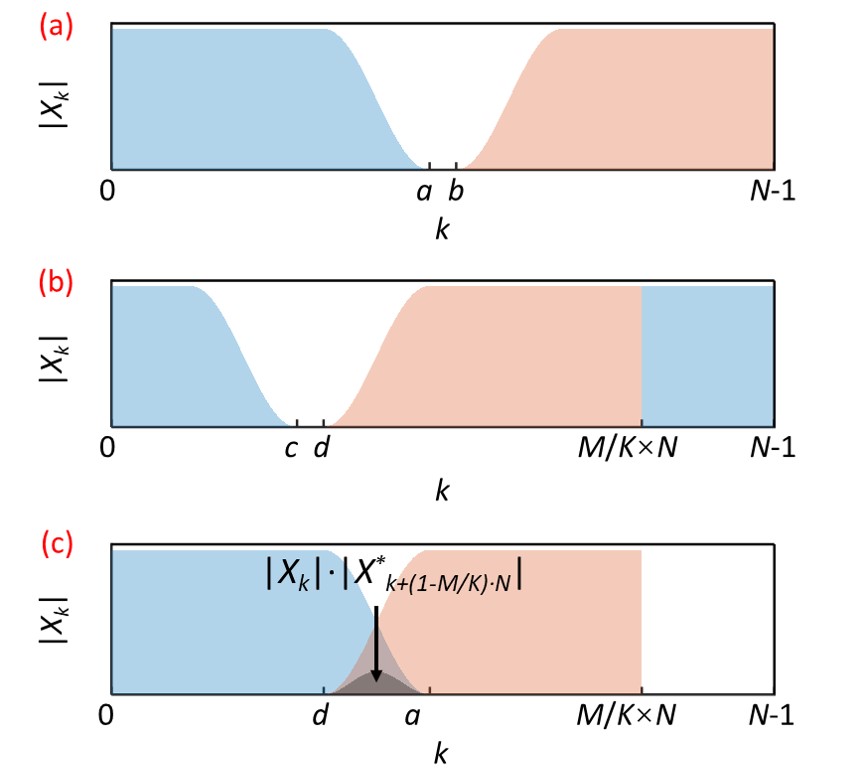}
\caption{(a) Magnitude spectrum of the received signal with a non-integer oversampling rate. (b) Magnitude spectrum of the received signal shifted by $M/K\times N$. (c) The overlap of two spectral parts of the spectrum.}
\label{FRAC_TR}
\end{figure}

The frequency-domain Godard TR can work at a non-integer oversampling rate close to 1 sps. The magnitude spectrum $X$ of the signal after match filtering with a roll-off factor $\beta$ is shown in Fig. \ref{FRAC_TR}(a). Fig. \ref{FRAC_TR}(b) shows the spectrum shifted by $(1 - M/K) \times N$. Fig. \ref{FRAC_TR}(c) shows the spectral parts $|X_k|$, $|X_{k + (1-M/K)N}|$, and their overlap. The multiplication between the spectral parts is zero except $k \in [d, a]$, which can be calculated as
\begin{equation}
d = \frac{(1-\beta)M}{2 K} N,
\end{equation}
\begin{equation}
a = \frac{(1+\beta)M}{2 K} N-1.
\end{equation}
Therefore, for the non-integer-oversampling Godard timing error detector, the timing error is estimated as \cite{josten2017modified, josten2017168}
\begin{equation}
\hat{\tau}_{\varepsilon}=\sum_{k=\frac{(1-\beta)M}{2 K} N}^{\frac{(1+\beta)M}{2 K} N-1} \operatorname{Im}\left[X_k \cdot X_{k+(1-M/K) N}^*\right]
\label{fractional_TR}
\end{equation}
where $\operatorname{Im}(\cdot)$ denotes the imaginary part of a complex. The DFT size $N$ should choose a value that allows the upper and lower bounds to be integers. Then the signal is transferred to the TD by inverse DFT (IDFT) and the subsequent algorithms are performed in the TD.

\subsection{Frame synchronization and fine FOE}
The frame synchronization can be implemented based on the stacked timing metric after resampling the non-integer-oversampled signal to 1 sps \cite{wang2023fast}. Then the synchronization position $p$ of the oversampled signal is obtained by
\begin{equation}
p = \lfloor p_1 \times K/M \rfloor
\label{sync_equation}
\end{equation}
where $\lfloor \cdot \rfloor$ denotes the round-down operation and $p_1$ is the synchronization position of the signal at 1 sps. The fractionally-spaced delay $D$ caused by the round-down operation can be calculated by
\begin{equation}
D = \left[\operatorname{mod}\left(p, K\right) - 1\right]/M
\end{equation}
where $\operatorname{mod}\left(p, K\right)$ denotes the remainder of $p$ divided by $K$.  

After the frame synchronization, a fine FOE is implemented to eliminate the residual frequency offset using a periodic training sequence, which can be represented as $\left[\boldsymbol{S}, \boldsymbol{S}, \boldsymbol{S}\right]$. The length of the sequence $\boldsymbol{S}$ is $L_s$. The fine FOE can be calculated as \cite{morelli1999improved}
\begin{equation}
\begin{aligned}
\Delta f & =\frac{R_s}{4 \pi L_s^2} \sum_{k=1}^{L_s} \arg \left[R_2(k) \times R_1^*(k)+R_3(k) \times R_2^*(k)\right] \\
& =\frac{R_s}{4 \pi L_s^2} \sum_{k=1}^{L_s} \arg \left[2\left|S(t)|_{t=k+\tau}\right|^2 \cdot \exp \left(j \frac{2 \pi L_s}{R_s} \Delta f(k)\right)\right]
\end{aligned}
\end{equation}
where $R_s$ is the baud rate and $\tau$ is a residual sampling phase. The $\arg(\cdot)$ represents the operation of taking the angle of a complex value and $(\cdot)^\mathbf{*}$ denotes the conjugate operation. $\left[\boldsymbol{R}_{1}, \boldsymbol{R}_{2}, \boldsymbol{R}_{3}\right]$ denotes the received training sequence $\left[\boldsymbol{S}, \boldsymbol{S}, \boldsymbol{S}\right]$ after TR and the resampling to 1 sps. It will have almost no influence on the fine FOE if there is a residual sampling phase $\tau$. 

\begin{figure}[!t]
\centering
\includegraphics[width=\linewidth]{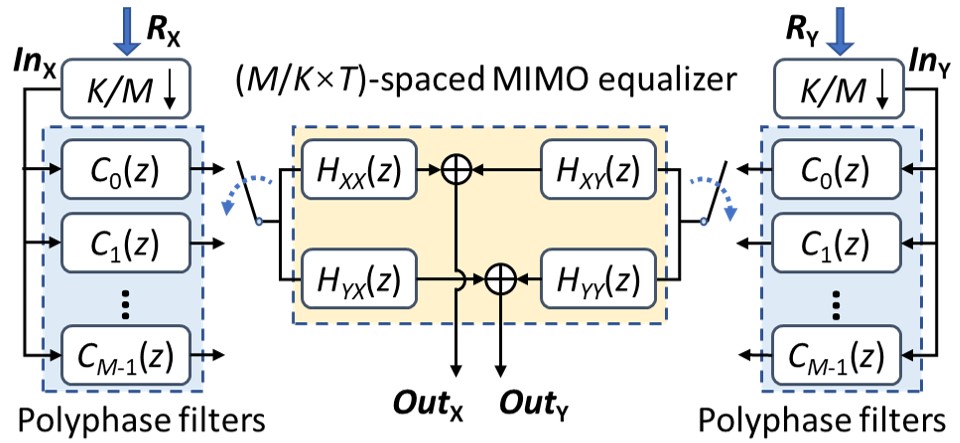}
\caption{Polyphase fractionally-spaced MIMO equalizer.}
\label{FRAC_EQ}
\end{figure}

\begin{figure*}[!t]
\centering
\includegraphics[width=\linewidth]{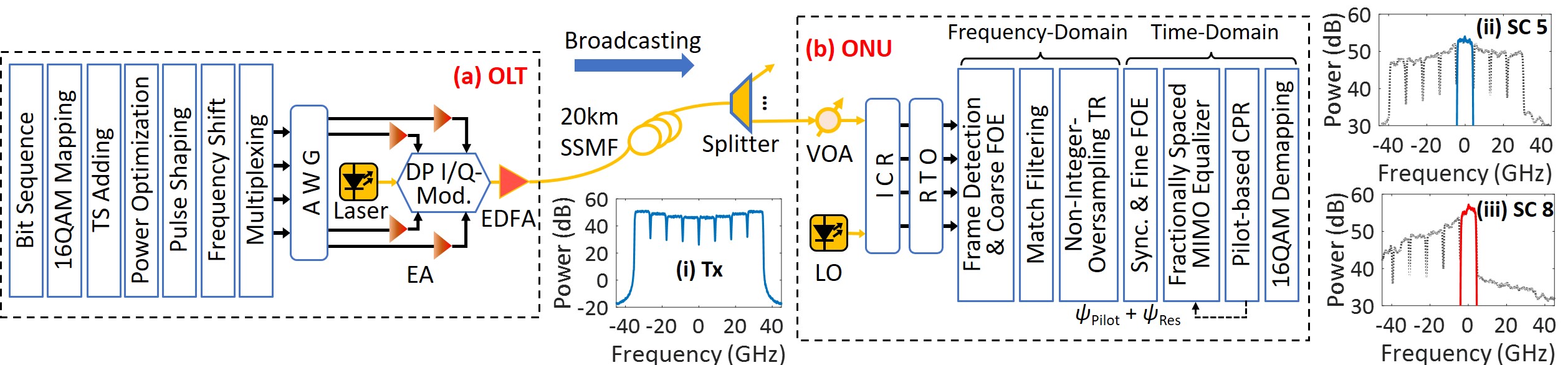}
\caption{Experimental setups of (a) OLT and (b) ONU at the $8$-Gbaud/SC$\times 8$-SCs DSCM-based coherent PON. Inset (i) is the electrical spectrum of the DSCM signal at the transmitter. Insets (ii) and (iii) are the received spectrum of the  allocated subcarrier 5 and subcarrier 8, respectively. DSCM signal is broadcast to ONU, which selects the allocated subcarrier by tuning the wavelength of LO.}
\label{EX}
\end{figure*}

\subsection{Polyphase fractionally-spaced MIMO equalizer}
The fractionally-spaced equalizer banks suffer from poor convergence because of $M$ independent fractionally-spaced equalizers \cite{shah2016low}. Fig. \ref{FRAC_EQ} shows the structure of the polyphase fractionally-spaced MIMO equalizer. The proposed fractionally-spaced MIMO equalizer consists of two polyphase filters and a $(M/K \times T)$-spaced MIMO equalizer. The polyphase filters consists of a bank of $M$ fractional delay filters $C_k(z)$ and $k = 0, 1, \cdots, M-1$. Polyphase filters compensate for the fractionally delay $D$ caused by frame synchronization and the fixed delay between the non-integer oversampled samples and the transmitted symbols \cite{shah2015low}. 

The received signal $R_{\mathrm{X,Y}}\left(n\right)$ is downsampled to 1 sps, which can be expressed as
\begin{equation}
In_{\mathrm{X,Y}}\left(n\right) = R_{\mathrm{X,Y}}\left(\lfloor n\times K/M\rfloor\right).
\end{equation}
The output of the polyphase filters using Lagrange's interpolation on the X or Y polarization can be expressed as 
\begin{equation}
p_{\mathrm{X,Y}}(n)=\sum_{i=-L}^{L} C_k(i) \cdot In_{\mathrm{X,Y}}\left(n-i\right)
\end{equation}
where $L$ is the half-tap length of the polyphase filters. $C_k(i)$ is the tap coefficients of the $i$-th tap of the $k$-th fractional delay filter.
$C_k(i)$ can be calculated as
\begin{equation}
C_k(i)=\prod_{j=-L / 2, j \neq i}^{L / 2} \frac{[\mu(k)]-j}{i-j}.
\end{equation}
$k$ can be calculated as
\begin{equation}
k =\operatorname{mod}(\lfloor n\times K/M\rfloor, K-1).
\end{equation}

The fractionally-spaced MIMO equalizer uses $\boldsymbol{p_{X,Y}}$ as the input to compensate for the channel impairments. The output $\boldsymbol{Out_{X,Y}}$ of the proposed fractionally-spaced MIMO equalizer is at the rate of $1/T$, which can be expressed as
\begin{equation}
Out_X(n) = \sum_{i=0}^{L_1-1} H_{XX}(i) \cdot p_{\mathrm{X}}(n - i) + H_{XY}(i) \cdot p_{\mathrm{Y}}(n - i),
\end{equation}
\begin{equation}
Out_Y(n) = \sum_{i=0}^{L_1-1} H_{YY}(i) \cdot p_{\mathrm{Y}}(n - i) + H_{YX}(i) \cdot p_{\mathrm{X}}(n - i),
\end{equation}
where $L_1$ is the tap length of the MIMO equalizer. The output of the equalizer is at 1 sps. Finally, CPR is implemented for the equalized signal to recover the signal. 

\section{Experimental setups}
\label{experimental_setups}

The experimental setups of an $8$-Gbaud/SC$\times 8$-SCs DSCM-based coherent PON are shown in Fig. \ref{EX}. At the transmitter side, the signal is generated by using the transmitter (Tx) DSP. Firstly, the bit sequence is mapped to the 16 quadrature amplitude modulation (16QAM) symbols. There are 9137 symbols in one frame at each subcarrier, including a 416-symbol training sequence (TS) and 273 pilot symbols. The pilot symbols are periodically inserted into the payload for a low-complexity CPR. The power of each subcarrier is optimized to balance the signal-to-noise ratio (SNR). Then the signal is pulse shaped by a root-raised cosine (RRC) filter with a 0.1 roll-off factor and shifted to the central frequency of each subcarrier. Finally, the DSCM signal is generated by subcarrier multiplexing and uploaded to a 96 GSa/s-arbitrary waveform generator (AWG). Inset (i) is the electrical spectrum of the DSCM signal. After being amplified by the electrical amplifiers (EAs), the analog signal is modulated by a dual-polarized in-phase/quadrature modulator (DP I/Q-Mod.) on an optical carrier @ $\sim$1545.123 nm. The output power of the modulator is about $-14$ dBm. An erbium-doped optical fiber amplifier (EDFA) is used as a booster amplifier to increase the launch optical power to 10dBm but causes about $-1.5$ dB penalty on receiver sensitivity. The optical signal is launched into a 20km standard single-mode fiber (SSMF). 

A variable optical attenuator is used at the receiver side to adjust the received optical power (ROP). A tunable ECL is used as the local oscillator (LO), which has a linewidth less than 100 kHz and $\sim$12.31 dBm output power. Insets (ii) and (iii) of Fig. \ref{EX} show the received spectrum of the allocated subcarrier 5 and subcarrier 8, respectively. Each of the eight subcarriers can be selected by tuning the wavelength of LO. An integrated coherent receiver (ICR) is used to mix the optical signal with LO and convert it into an electrical signal and a real-time oscilloscope (RTO, Keysight UXR0594AP) digitizes the signal. Then the digital signal is resampled to 9/8-sps, 5/4-sps, and 2 sps, respectively. Finally, the received signal is recovered by the receiver-end DSP, including frame detection, coarse FOE, match filtering, and TR performed in the FD, while the frame synchronization, fine FOE, equalization, and CPR in the TD. The line rate of the DSCM-based coherent PON is 512 Gb/s (i.e. $8$-Gbaud/SC$\times 8$-SCs $\times 4$ bits/symbol/polarization $\times$ 2 polarization $= 512$ Gb/s). When the 20\% soft-decision forward error correction (SD-FEC) is considered, the net rate is approximately 400 Gb/s.

After the frame detection, coarse FOE is implemented. Then the signal is filtered by an RRC match filter. TR is implemented based on the non-integer-oversampling Godard timing error detector. After the frame synchronization and fine FOE, a fractionally-spaced MIMO equalizer is used to compensate for the channel impairments. The output signal of the equalizer is at 1 sps. The tap coefficients of the fractionally-spaced MIMO equalizer are converged using the least mean square (LMS) algorithm. The phase noise $\psi_{\text {Pilot}}$ is estimated using the periodically inserted pilot symbols and the residual phase noise $\psi_{\text {Res}}$ is estimated using a $(2Q+1)$-tap average filter at 1 sps. Then the total phase noise $(\psi_{\text {Pilot}} + \psi_{\text {Res}})$ is compensated for the baud-rate equalized signal and fed to the decision-directed LMS algorithm to track the tap coefficients, which requires CPR before the calculation of the error function. Finally, the recovered 16QAM symbols are de-mapped to the bits.

\begin{figure}[!t]
\centering
\includegraphics[width=0.945\linewidth]{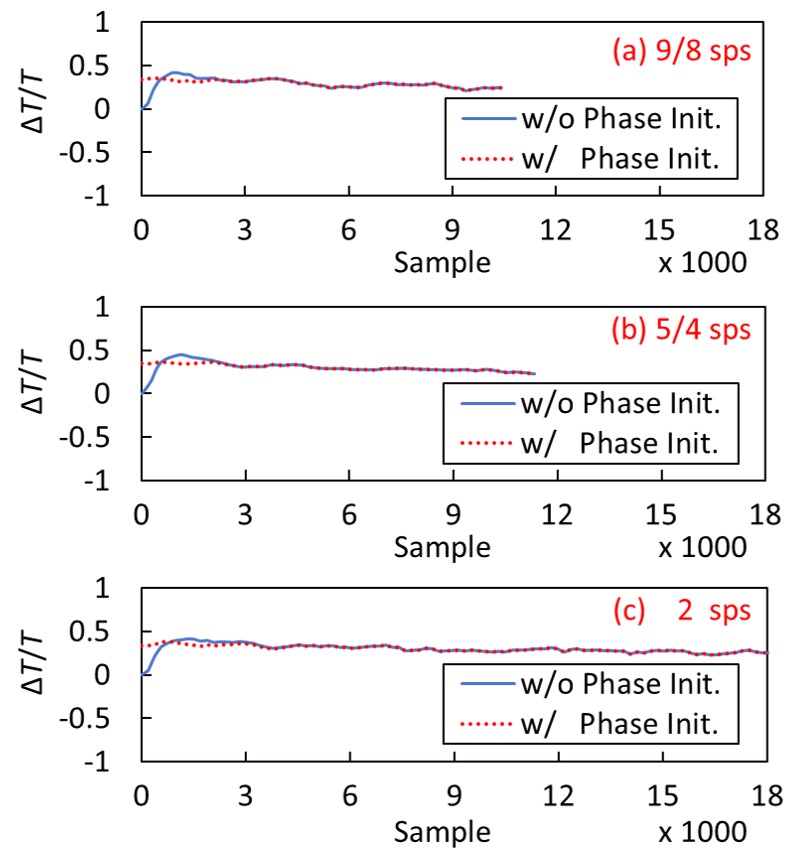}
\caption{The estimated timing error without and with the sampling phase initialization of the TR at (a) 9/8-sps, (b) 5/4-sps, and (c) 2-sps oversampling rate, respectively.}
\label{ted}
\end{figure}

\section{Experimental results and discussions}\label{ex_results}

To ensure the upper and lower bounds of Eq. (\ref{fractional_TR}) to be integers, the DFT size $N$ is 252, 250, and 256 for the 9/8-sps, 5/4-sps, and 2-sps oversampling TR, respectively. The overlap size is set to 54, 50, and 64, which results in the overlap rate $\eta$ of 1.27, 1.25, and 1.33, respectively. After the detection of the training sequence with the two symmetric frequency tones at $\pm 1/2T + \Delta f$, the coarse FOE is implemented to compensate for the offset $\Delta f$ by a cyclic shift of the frequency components. After match filtering, TR is executed. Fig. \ref{ted} shows the estimated timing error without and with the sampling phase initialization of the TR at (a) 9/8-sps, (b) 5/4-sps, and (c) 2-sps oversampling rates, respectively. Godard TR algorithm can estimate the sampling phase $\Delta T$ accurately. An initial sampling phase is estimated using the same training sequence for coarse FOE, which can accelerate the convergence of TR \cite{wang2023fast}. The sampling phase initialization ensures the validity of frame synchronization, which is performed subsequently. Under the same symbol length, the lower the oversampling rate, the fewer samples need to be processed. Therefore, the non-integer-oversampling TR can reduce the high computational complexity for coherent PON.

\begin{figure}[!t]
\centering
\includegraphics[width=0.945\linewidth]{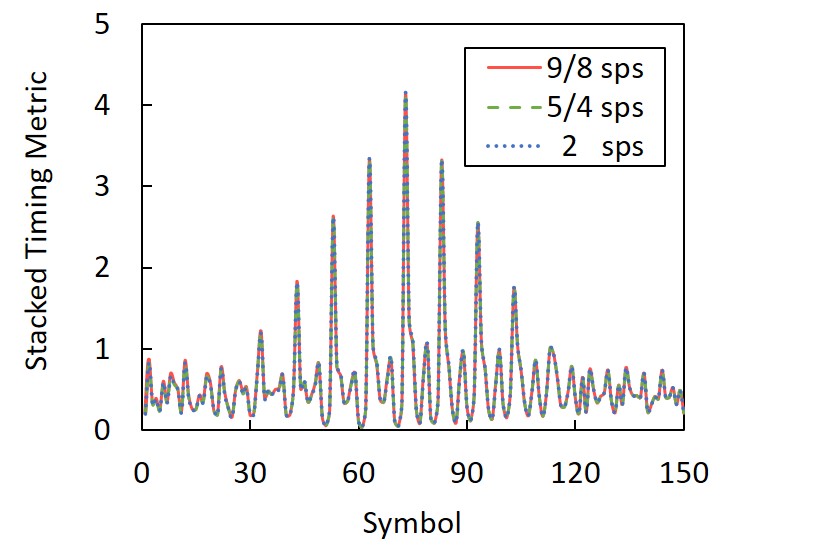}
\caption{The stacked timing metric calculated at 1 sps for frame synchronization after the TR at 9/8-sps, 5/4-sps, and 2-sps oversampling rate, respectively.}
\label{sync}
\end{figure}

The training sequence for frame synchronization follows the training sequence for frame detection and coarse FOE. As a result, after frame detection and coarse FOE, the approximate position of the training sequence can be obtained for frame synchronization. Therefore, only the subsequent $L_f$ ($L_f = 100$) symbols are required by the sliding window for frame synchronization. The training sequence is a periodic sequence with a length of 64 symbols. Frame synchronization is realized by calculating the stacked timing metric after resampling all the oversampled signals to 1 sps. There are nine equally spaced peaks in the stacked timing metric as shown in Fig. \ref{sync}. The fifth peak is the highest peak without a plateau, which indicates the accurate synchronization position $p_1$. After resampling the 9/8-sps, 5/4-sps, and 2-sps oversampled signals to 1 sps, the stacked timing metrics are almost identical. Since the sampling phase error can affect the frame synchronization, the identical timing metrics also mean that the non-integer-oversampling TR is effective. Then the synchronization position $p$ of the oversampled signal is calculated as the Eq. (\ref{sync_equation}).

\begin{figure}[!t]
\centering
\includegraphics[width=0.945\linewidth]{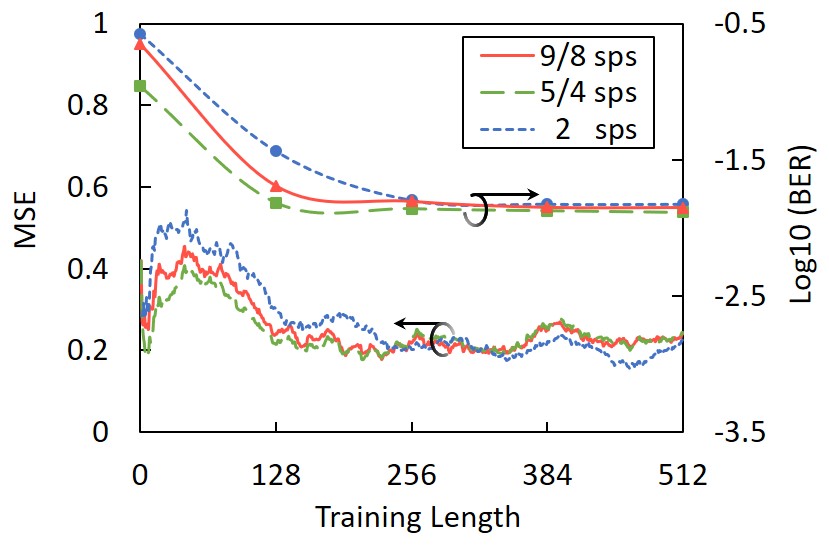}
\caption{MSE and BER using fractionally-spaced MIMO equalizers with 9/8-sps, 5/4-sps, and 2-sps oversampling rates, respectively.}
\label{MSE}
\end{figure}

\begin{figure*}[!t]
\centering
\includegraphics[width=\linewidth]{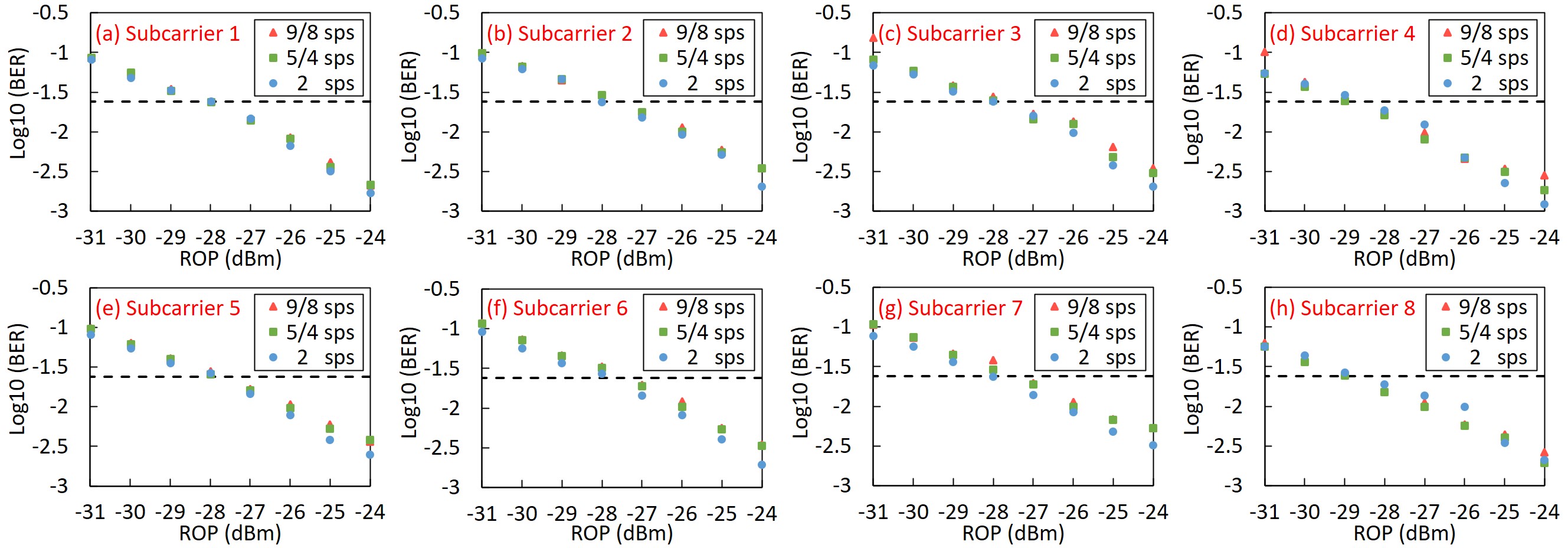}
\caption{BER performance versus ROP of the coherent PON using 9/8-sps, 5/4-sps, and 2-sps oversampling DSP from subcarrier 1 to subcarrier 8, respectively. The dashed line represents the 20\% SD-FEC limit.}
\label{BER}
\end{figure*}
\begin{table*}[!t]
	\centering
	\begin{threeparttable}[b]
	\renewcommand\arraystretch{1.2}
	\caption{Computational complexity per symbol in terms of complex-valued multiplication (Mul.) and addition (Add.).}
	\begin{tabular}{cccccccc}
\Xhline{1px}
	     &\begin{tabular}[c]{@{}c@{}}{DFT/IFFT}\end{tabular} & \begin{tabular}[c]{@{}c@{}}Match\\Filter\end{tabular} & \begin{tabular}[c]{@{}c@{}}Timing\\Recovery\end{tabular} & \begin{tabular}[c]{@{}c@{}}Frame\\Synchronization\end{tabular} &\begin{tabular}[c]{@{}c@{}}Fine FOE\end{tabular} & \begin{tabular}[c]{@{}c@{}}MIMO \\Equalization\end{tabular} &\begin{tabular}[c]{@{}c@{}}Pilot-based \\CPR\end{tabular}\\ 
\hline
		Mul. &$\eta NK/M$ &$\eta K/M$ & $\eta(\beta + K/M)$ & $(4L_s + 4)L_f/F$ & $(4L_s + 1)/F + K/M$  &$(2L + 1)$ + $2L_1$ & $P/F + 2Q+1$ \vspace{2pt} \\ 
		Add. &$\eta(N-1)K/M$ &0 & $\eta(\beta - K/M/N)$ & $(3L_s + 2)L_f/F$ &$(4L_s - 1)/F$  &$2L + 2(L_1 - 1)$ &$2Q$\\ 
\Xhline{1px}
	\end{tabular}
       \footnotesize
       $\eta$: overlap rate, $N$: DFT/IDFT size, $K/M$: oversampling rate, $\beta$: roll-off factor, $L_s$: the size of sliding window, $L_f$: length of symbols for frame synchronization, $F$: symbol number at one data frame, $L$: the half-tap length of the polyphase filter, $L_1$: tap length of the MIMO equalizer, $P$: number of pilot symbols, $Q$: half-length of the average filter.
     \label{tab}
     \end{threeparttable}
\end{table*}

After frame synchronization and fine FOE, the fractionally-spaced MIMO equalizer is trained using the LMS algorithm. The half-tap length $L$ of the polyphase filter is 14 and 7 for DSP with 9/8-sps and 5/4-sps oversampling rates, respectively. For DSP with an integer-oversampling rate, the polyphase filters are not required, which means that the half-tap length $L$ of the polyphase filter is zero. The tap length $L_1$ of MIMO equalizer for the 9/8-sps, 5/4-sps, and 2-sps oversampling rates are optimized to 9, 9, and 11, respectively. Fig. \ref{MSE} shows the mean square error (MSE) of fractionally-spaced MIMO equalizers and the bit-error rate (BER) using DSP with 9/8-sps, 5/4-sps, and 2-sps oversampling rates, respectively. Three MSEs can achieve convergence within approximately 256 training symbols. After equalization, the signal is downsampled to the baud rate. CPR is implemented at 1 sps using pilot symbols and an average filter with the half-length $Q$ of $7$. BER can be reflected by MSE curves. When the training length is approximately 256, the BER after the non-integer-oversampling DSP does not decrease significantly, which is close to the BER of the 2-sps oversampling DSP. Therefore, there is almost no penalty on the BER performance. 

Figure \ref{BER} shows the BER performance of all the subcarriers versus ROP of the 400-Gb/s-net-rate coherent PON using 9/8-sps, 5/4-sps, and 2-sps oversampling DSP from subcarrier 1 to subcarrier 8, respectively. The dashed line represents the 20\% SD-FEC limit. Due to the low baud rate of each subcarrier in TFDMA-coherent PON using DSCM, the relatively low dispersion caused by the 20-km SSMF for each 8-Gbaud subcarrier can be compensated by only the MIMO equalizer. Therefore, no additional chromatic dispersion compensation module is required, which can greatly alleviate the DSP burdens for ONUs. BERs of all the eight subcarriers of the DSCM signal using DSP working at 9/8-sps, 5/4-sps, and 2-sps oversampling rates can achieve below the 20\% SD-FEC limit when the ROP is no less than $-27$ dBm. Using the proposed non-integer-oversampling DSP, BER can be close to that of the 2-sps oversampling DSP with almost no penalty on the receiver sensitivity.

The computational complexity per symbol in terms of complex-valued multiplication (Mul.) and addition (Add.) of the non-integer-oversampling DSP is listed in Table \ref{tab}. The oversampling sampling rate $K/M$ leads to a different total number of samples to be processed. Therefore, the main difference in computational complexity of the DSP with different over-sampling rates lies in the algorithms related to the oversampling sampling rate $K/M$, including all FD algorithms and the TD equalization. However, frame synchronization, fine FOE, and CPR are all implemented in the TD at 1 sps, which leads to the same computational complexity for those parts of DSP with different sampling rates. The coarse FOE only requires a cyclic shift of the frequency components but no multiplication and addition. DFT and IDFT would account for the major part of the computational complexity of the non-integer-oversampling DSP. Therefore, reducing the oversampling rate $K/M$ is an important way to lower the computational complexity of DSP if the DFT size $N$ and the overlap rate $\eta$ are almost the same for all the oversampling rates.

\begin{figure}[!t]
\centering
\includegraphics[width=0.945\linewidth]{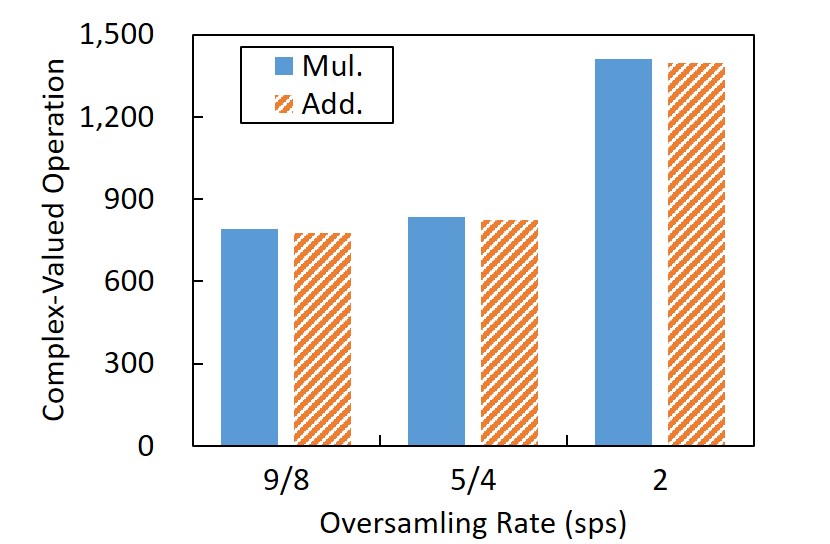}
\caption{Computational complexity per symbol of the DSP with 9/8-sps, 5/4-sps, and 2-sps oversampling rates, respectively.}
\label{complexity}
\end{figure}

The exact computational complexity per symbol (i.e., the total number of complex-valued multiplication and addition) of the DSP with 9/8-sps, 5/4-sps, and 2-sps oversampling rates is shown in Fig. \ref{complexity}. Due to the reduction of the total number of processing samples, the overall computational complexity per symbol of the non-integer-oversampling DSP can be greatly reduced. Compared to the DSP working at a 2-sps oversampling rate, the complex-valued multiplication of the DSP working at the 9/8-sps and 5/4-sps oversampling rate is reduced by 44.04\% and 40.78\%, respectively. Moreover, the complex-valued addition is reduced by 44.26\% and 41.01\%, respectively. Therefore, the non-integer-oversampling DSP can reduce the computational complexity for coherent PON. The high sampling rate requirement is also lowered, which will reduce the cost and is highly preferred for ONUs. It is worth mentioning that although the 9/8-sps oversampling DSP has a lower computational complexity compared to the 5/4-sps oversampling DSP, it needs more fractional delay filters in the polyphase filters, which leads to a more complex structure. 

\section{Conclusions}
\label{conclusions}
In this paper, we propose a non-integer-oversampling DSP to reduce the computational complexity for TFDMA-coherent PON. The proposed DSP mainly includes the non-integer-oversampling FOE, TR, frame synchronization, and polyphase fractionally-spaced MIMO equalizer. Compared to the 2-sps oversampling rate, the computational complexity of the proposed algorithm working at 9/8-sps and 5/4-sps oversampling rates can be reduced by 44.04\% and 40.78\%, respectively. Moreover, the experimental results of a 400-Gb/s-net-rate coherent PON using DSCM show that there is almost no penalty on the receiver sensitivity. The proposed non-integer-oversampling DSP can reduce the computational complexity and lower the high sampling rate requirement of ADC, which is preferred in coherent PON. To further reduce the computational complexity of DSP for coherent PON, the simplified non-integer-oversampling DSP is worth studying. Therefore, the proposed non-integer-oversampling DSP shows great potential in the future coherent PON.

\ifCLASSOPTIONcaptionsoff
  \newpage
\fi
\footnotesize
\bibliographystyle{IEEEtran}
\bibliography{sample}
\end{document}